\documentclass[twocolumn,showpacs,superscriptaddress,
preprintnumbers,floatfix]{revtex4}
\usepackage{graphicx,epsf,amssymb,amsmath,latexsym}
\newcommand{\be}{\begin{eqnarray}}
\newcommand{\ee}{\end{eqnarray}}

\renewcommand{\d}{\mbox{${\rm d}$}}

\begin{document}
\title{\bf Improved WKB analysis of cosmological perturbations}
\author{Roberto Casadio}
\email{Roberto.Casadio@bo.infn.it}
\affiliation{Dipartimento di Fisica, Universit\`a di Bologna and I.N.F.N.,
Sezione di Bologna, via Irnerio~46, 40126 Bologna, Italy}
\author{Fabio Finelli}
\email{finelli@bo.iasf.cnr.it}
\affiliation{IASF/INAF, Istituto di Astrofisica Spaziale e Fisica
Cosmica, Istituto Nazionale di Astrofisica, Sezione di Bologna,
via Gobetti~101, 40129 Bologna, Italy.}
\author{Mattia Luzzi}
\email{Mattia.Luzzi@bo.infn.it}
\affiliation{Dipartimento di Fisica, Universit\`a di Bologna and I.N.F.N.,
Sezione di Bologna, via Irnerio~46, 40126 Bologna, Italy}
\author{Giovanni Venturi}
\email{armitage@bo.infn.it}
\affiliation{Dipartimento di Fisica, Universit\`a di Bologna and I.N.F.N.,
Sezione di Bologna, via Irnerio~46, 40126 Bologna, Italy}
\begin{abstract}
Improved Wentzel-Kramers-Brillouin (WKB)-type approximations
are presented in order to study cosmological perturbations
beyond the lowest order.
Our methods are based on functions which approximate the true
perturbation modes over the complete range of the independent
(Langer) variable, from sub-horizon to super-horizon scales,
and include the region near the turning point.
We employ both a perturbative Green's function technique
and an adiabatic (or ``semiclassical'') expansion (for a
linear turning point) in order to compute higher order
corrections.
Improved general expressions for the WKB scalar and tensor
power spectra are derived for both techniques.
We test our methods on the benchmark of power-law inflation,
which allows comparison with exact expressions for the
perturbations, and find that the next-to-leading order
adiabatic expansion yields the amplitude of the power spectra
with excellent accuracy, whereas the next-to-leading order with
the perturbative Green's function method does not improve
the leading order result significantly.
However, in more general cases, either or both methods may
be useful.
\end{abstract}
\pacs{98.80.Cq, 98.80.-k}
\maketitle
\section{Introduction}
\label{intro}
Anisotropies in the cosmic microwave background (CMB)
radiation and inhomogeneities in the large scale structures
of the Universe have nowadays become a fundamental tool
to study the early universe.
This trend will persist in the next years due to the data
releases of Wilkinson~Microwave~Anisotropy~Probe~\cite{wmap},
Sloan~Digital~Sky~Survey~\cite{sdss} and the 
launch of the Planck mission~\cite{planck}. 
\par
In a primordial inflationary phase, matter and space-time
fluctuations are generated which inherit characteristics from
the particular inflationary model studied (for an introduction
and review see, e.g.~Ref.~\cite{infla}).
It is therefore very important to be able to determine the power
spectra of cosmological perturbations for a large variety
of different inflationary models, so that comparison with
available and future data will tell us which models
satisfactorily represent the time evolution of
the Universe.
Since exact solutions are not available for cosmological
perturbations in general inflationary models (except for
the case of an exponential potential~\cite{lythstewart})
approximation methods are very welcome.
The method of approximation most used is the slow-rollover
approximation, introduced by Stewart and Lyth~\cite{SL}
(see Ref.~\cite{LLKCTBA} for a review), which
was inspired by power-law inflation.
\par
It is also worthwhile testing different approximation schemes.
In this paper we present Wentzel-Kramers-Brillouin (WKB)-type
approaches which improve the proposals of
Refs.~\cite{martin_schwarz,H_MP}
(see also Refs.~\cite{mukh,mukh_thory,wang_mukh,SGong,TW,FMVV_II}
for other approximations).
For power-law inflation, the usual WKB method to lowest order
reproduces exactly the spectral indices of perturbations,
but the amplitudes are still rendered poorly.
Given the importance of the consistency equation between
the ratio of tensor to scalar amplitudes and the spectral index
of tensor modes for single Klein-Gordon scalar field
inflation~\cite{infla}, we believe that the prediction on the
amplitudes should also be improved, as is done
in Ref.~\cite{HHHJM} by means of the so-called
``uniform approximation'' to next-to-leading order
(for the details on the formalism, see Ref.~\cite{olver}).
\par
Let us begin by recalling the Robertson-Walker metric in
conformal time $\eta$,
\be
\d s^2=a^2(\eta)\,\left[-\d\eta^2+\d\vec x\cdot \d\vec x\right]
\ ,
\ee
where $a$ is the scale factor of the Universe.
Scalar (density) and tensor (gravitational wave) fluctuations
are determined on this background by two functions, respectively
denoted as $\mu=\mu_{\rm S}:= a\,Q$ ($Q$ is the Mukhanov
variable~\cite{mukh}) and $\mu=\mu_{\rm T}:= a\,h$
($h$ is the amplitude of the two polarizations
of gravitational waves)~\cite{gris,staro},
which satisfy a one-dimensional Schr\"odinger-like equation
\be
\left[\frac{\d^2}{\d\eta^2}
+k^2-U(\eta)\right]\,\mu
=
\left[\frac{{\d}^2}{{\d}\eta^2}+\Omega^2(k,\eta)\right]
\,\mu=0
\ ,
\label{osci}
\ee
where $k$ is the wave-number.
The time-dependent frequency is given by
\be
\Omega^2(k,\eta)=k^2-\frac{z''}{z}
\ ,
\label{freq}
\ee
where $z=z_{\rm S}:=a\,\phi'/{\cal H}$ for scalar ($\phi$
is the homogenous inflaton) and $z=z_{\rm T}:=a$ for tensor
perturbations, primes denote derivatives with respect to $\eta$,
and ${\cal H}$ ($:= a'/a$) is the conformal Hubble parameter.
Eq.~(\ref{osci}) must be solved together with the condition that
the modes are initially plane waves for wavelengths much
shorter than the Hubble radius,
\be
\lim_{\frac{k}{a\,H}\rightarrow +\infty} \mu(k,\eta)
=d\,\frac{{\rm e}^{-i\,k\,\eta}}{\sqrt{2\,k}}
\ ,
\label{init_cond_on_mu}
\ee
where $d=d_{\rm S}:=1$ and
$d=d_{\rm T}:=\sqrt{8\,\pi}/m_{\rm Pl}$
for scalar and tensor perturbations ($m_{\rm Pl}$ is
the Planck mass).
The dimensionless power spectra of scalar and tensor
fluctuations are then given by
\begin{subequations}
\be
\mathcal{P}_{\zeta}=
\displaystyle\frac{k^{3}}{2\,\pi^{2}}\,
\left|\frac{\mu_{\rm S}}{z_{\rm S}}\right|^{2}
\ ,
\ \
\mathcal{P}_{h}=
\displaystyle\frac{4\,k^{3}}{\pi^{2}}\,
\left|\frac{\mu_{\rm T}}{z_{\rm T}}\right|^{2}
\ .
\label{spectra_def}
\ee
The spectral indices and their runnings are defined at an arbitrary
pivot scale $k_*$ as
\be
n_{\rm S}-1:=
\left.\displaystyle\frac{\d\ln \mathcal{P}_{\zeta}}
{\d\ln k}\right|_{k=k_{*}}
\ ,
\ \
n_{\rm T}:=
\left.\displaystyle\frac{\d\ln \mathcal{P}_{h}}
{\d\ln k}\right|_{k=k_{*}}
\ ,
\label{n_def}
\ee
and
\be
\alpha_{\rm S}:=\left.
\frac{\d^{2}\ln\mathcal{P}_{\zeta}}
{(\d\ln k)^{2}}\right|_{k=k_{*}}
\ ,
\ \
\alpha_{\rm T}:=\left.
\frac{\d^{2}\ln \mathcal{P}_{h}}
{(\d\ln k)^{2}}\right|_{k=k_{*}}
\ .
\label{alpha_def}
\ee
\end{subequations}
\par
The evolution of the Universe is usually described by means of
a set of flow equations~\cite{martin_schwarz,terrero,tesi}.
The zero horizon flow function is defined by
\be
\epsilon_{0}:=\frac{H(N_{i})}{H(N)}
\ ,
\label{eps_0}
\ee
where $H:=\dot{a}/a$ is the Hubble rate and dots denote derivatives
with respect to cosmic time $\d t=a(\eta)\,\d\eta$.
In this expression, $N$ is the number of e-folds, $N:=\ln (a/a_i)$
[where $a_i=a(\eta_i)$] after the arbitrary initial time $t_i=t_i(\eta_i)$.
The hierarchy of horizon flow functions is then defined according to
\be
\epsilon_{n+1}:=
\frac{\d\ln |\epsilon_{n}|}{\d N}
\ ,
\quad\quad
n\geq 0
\ ,
\label{hor_flo_fun}
\ee
and we recall that inflation takes place for $\epsilon_1 < 1$.
\par
The effective potentials $U=U_{\rm S}$ and $U=U_{\rm T}$
can now be expressed in terms of $\epsilon_1$,
$\epsilon_2$, and $\epsilon_3$ only as
\be
&&
\frac{U_{\rm T}(\eta)}{a^{2}\,H^{2}}=2-\epsilon_{1}
\nonumber
\\
\label{pot}
\\
&&
\frac{U_{\rm S}(\eta)}{a^{2}\,H^{2}}
=
2-\epsilon_{1}+\frac{3}{2}\,\epsilon_{2}
-\frac{1}{2}\,\epsilon_{1}\,\epsilon_{2}
+\frac{1}{4}\,\epsilon_{2}^{2}
+\frac{1}{2}\,\epsilon_{2}\,\epsilon_{3}
\ ,
\nonumber
\ee
and Eq.~(\ref{osci}) can be solved exactly for $\epsilon_1$
constant in time, which is the case of an exponential potential
for the inflaton.
When the $\epsilon_n$ have an arbitrary time dependence,
approximate methods are necessary.
\par
A WKB analysis of this problem has been recently presented
in Ref.~\cite{martin_schwarz} to first order in the adiabatic (or
``semiclassical'') expansion (see next~Section and Ref.~\cite{birrel}
for precise definitions), in which the Langer transformation~\cite{langer}
\be
x:=\ln\left(\frac{k}{H\,a}\right)
\ ,\quad
\chi:=(1-\epsilon_1)^{1/2}\,{\rm e}^{-x/2}\,\mu
\ ,
\label{transf}
\ee
introduced in Ref.~\cite{wang_mukh} is used~\footnote{Note
that we define $x$ with the opposite sign with respect to that in
Ref.~\cite{martin_schwarz}.
This of course implies that our $x_i\to+\infty$ and $x_f\to-\infty$,
whereas $x_i\to-\infty$ and $x_f\to+\infty$ in
Ref.~\cite{martin_schwarz}.}
in order to improve the accuracy.
Such a transformation brings Eq.~(\ref{osci}) into the form
\be
\left[\frac{{\d}^2}{{\d} x^2}+\omega^2(x)\right]
\,\chi=0
\ ,
\label{new_eq}
\ee
where the (new) frequency $\omega(x)$ in general vanishes at
the classical ``turning point'' $x=x_*$.
On sub-horizon scales ($x\gg x_*$), the mode $\chi$ oscillates
($\omega\simeq k$) and the standard WKB approximation can be
applied without difficulty.
On super-horizon scales ($x\ll x_*$), the perturbations
exponentially decay or grow and the standard WKB approximation
can again be applied.
One then matches the two approximate solutions at $x=x_*$
to obtain the super-horizon amplitudes at $x_f\ll x_*$
which determine the CMB spectra.
This procedure opens up the possibility of further improving
the knowledge of the spectra by including subsequent adiabatic
orders.
\par
However, a straightforward generalization to higher adiabatic
orders is seriously hindered by the lack of a general
prescription for matching the two WKB branches with sufficient
accuracy.
This problem was thoroughly discussed in
Refs.~\cite{langer34,langer},
where it was also suggested to replace the standard
(plane wave-like) WKB functions with Bessel's functions.
The latter indeed remain good approximations at the turning
point, whereas the former require the matching with yet
another particular solution at some (unspecified) points
both to the left and right of $x=x_*$.
We shall follow the proposal of Ref.~\cite{langer} for a
linear turning point and show that the inclusion of higher
adiabatic orders indeed improve the results, contrarily
to what was found in Ref.~\cite{hunt}.
We shall also introduce a new (perturbative) expansion
for general turning point, which makes use of the Green's
function technique, and compare the corresponding corrections
for the case of power law inflation.
\par
The article is organized as follows:
In Section~\ref{WKB_power_spectra}, we shall review the
application to inflationary cosmological perturbations
of the standard WKB approximation to leading adiabatic
order~\cite{martin_schwarz} and its shortcomings.
In Section~\ref{schiffmethod}, we shall describe in full
generality a method of improving the results to
higher orders, both by the perturbative expansion discussed
in detail in Section~\ref{pert} and the adiabatic expansion 
reviewed in Section~\ref{adiab}.
In Section~\ref{Power-law inflation}, we shall apply the improved WKB
approximation to power-law inflation which, being exactly solvable,
will allow us to test the improved WKB predictions against exact
solutions.
Finally, we shall give our conclusions and observations
in Section~\ref{conc}.
\section{Standard WKB approximation}
\label{WKB_power_spectra}
In this Section we begin by recalling that the straightforward
application of the standard WKB method leads to a poor
approximation and that the results can be improved to first adiabatic
order by means of the transformation~(\ref{transf}), as was shown in
Ref.~\cite{martin_schwarz}.
However, the extension to higher orders is more subtle, as we shall
discuss in some detail.
\par
The WKB approximation is defined by first introducing a ``small''
parameter $\delta>0$ in order to perform the formal adiabatic
(or semiclassical) expansion of the mode
functions~\cite{birrel}~\footnote{Let us briefly recall that such
an expansion is called adiabatic when $\sqrt{\delta}$ is the
inverse of a (typically very large) time over which the system
evolves slowly, and semiclassical when $\sqrt{\delta}\sim\hbar$.}.
Consequently, Eq.~(\ref{osci}) is formally replaced by
\be
\left[\delta\,\frac{\d^2}{\d\eta^2}+\Omega^2(k,\eta)\right]\,
\mu=0
\ ,
\label{eq_wkb_exp}
\ee
and, of  course, the limit $\delta\to 1$ must be taken at the
end of the computation.
We then denote the leading order term in such an expansion
of $\mu$ by
\be
\mu_{{\rm WKB}}(k,\eta )=
\frac{{\rm e}^{\pm\frac{i}{\sqrt{\delta}}\,\int^{\eta }\Omega(k,\tau)
{\rm d}\tau}}{\sqrt{2\,\Omega(k,\eta)}}
\ ,
\label{wkb_sol}
\ee
which satisfies the following differential equation
\be
\left[\delta\,\frac{\d^2}{\d \eta^2}
+\Omega^2(k,\eta)-\delta\,Q_\Omega(k,\eta)\right]\,
\mu_{{\rm WKB}}=0
\ ,
\label{eq_wkb}
\ee
where
\be
Q_\Omega(k,\eta):=
\frac{3}{4}\,\frac{(\Omega')^2}{\Omega^2}
-\frac{\Omega''}{2\,\Omega}
\ .
\label{Q}
\ee
We now observe that $\mu_{{\rm WKB}}$ is the exact solution of
Eq.~(\ref{eq_wkb_exp}) in the ``adiabatic limit'' $\delta\to 0$,
and is expected to be a good approximation to the exact $\mu$
(in the limit $\delta\to 1$) if 
\be
\Delta:=
\left\vert\frac{Q_\Omega}{\Omega^2}\right\vert
\ll 1
\ .
\label{WKB_cond}
\ee
\par
On using the potentials in Eqs.~(\ref{pot}) in terms of
$a$ and $H$, one finds
\be
\Delta_{\rm S,T}=
\left\{ \begin{array}{ll}
\displaystyle\left(\frac{a\,H}{k}\right)^{4}
\mathcal{O}(\epsilon_{n})\sim 0
&
\textrm{sub-horizon scales}
\\
&
\\
\displaystyle\frac{1}{8}+\mathcal{O}(\epsilon_{n})\
&
\textrm{super-horizon scales.}
\end{array}
\right.
\label{cond_prechange}
\\
\nonumber
\ee
The result for super-horizon scales suggests that a change of
variable and function are necessary in order to obtain better
accuracy.
\par
In fact, on employing the transformation (\ref{transf}), one finds
(for $\delta=1$) the new equations of motion (\ref{new_eq}),
which explicitly read
\begin{widetext}
\begin{subequations}
\be
&&
\frac{\d^2\chi_{\rm S}(x)}{\d x^2}
+\left[\frac{{\rm e}^{2\,x}}{(1-\epsilon_1)^2}
-\frac{1}{4}\,\left(\frac{3-\epsilon_1}{1-\epsilon_1}\right)^2
-\frac{(3-2\,\epsilon_1)\,\epsilon_2}{2\,(1-\epsilon_1)^2}
-\frac{(1-2\,\epsilon_1)\,\epsilon_2\,\epsilon_3}{2\,(1-\epsilon_1)^3}
-\frac{(1-4\,\epsilon_1)\,\epsilon_2^2}{4\,(1-\epsilon_1)^4} \right]\,
\chi_{\rm S}(x)=0
\label{eqS_x}
\\
\nonumber
\\
&&
\frac{\d^2 \chi_{\rm T}(x)}{\d x^2}
+\left[\frac{{\rm e}^{2\,x}}{(1-\epsilon_1)^2}
-\frac{1}{4}\,\left(\frac{3-\epsilon_1}{1-\epsilon_1}\right)^2
+\frac{\epsilon_1\,\epsilon_2}{2\,(1-\epsilon_1)^2}
+\frac{\epsilon_1\,\epsilon_2\,\epsilon_3}{2\,(1-\epsilon_1)^3}
+\frac{(2+\epsilon_1)\,\epsilon_1\,\epsilon_2^2}{4\,(1-\epsilon_1)^4}
\right]\,
\chi_{\rm T}(x)=0
\ .
\label{eqT_x}
\ee
%
\end{subequations}
From $\omega_{\rm S}^2(x)$ and $\omega_{\rm T}^2(x)$, respectively
given by the expressions in the square brackets of
Eqs.~(\ref{eqS_x}) and~(\ref{eqT_x}), we now obtain
%
\be
\left\{\begin{array}{ll}
\Delta_{\rm S,T}
=e^{-2\,x}\mathcal{O}(\epsilon_{n})\sim 0
&
\quad
\textrm{sub-horizon scales}
\\
\\
\Delta_{\rm S}
=\displaystyle\frac{4}{27}\,
\left(\epsilon_{1}\,\epsilon_{2}^2
+\epsilon_{1}\,\epsilon_{2}\,\epsilon_{3}+
\frac{\epsilon_{2}\,\epsilon_{3}^2}{2}
+\frac{\epsilon_{2}\,\epsilon_{3}\,\epsilon_{4}}{2}\right)
+\mathcal{O}(\epsilon_{n}^{4})
&
\quad
\textrm{super-horizon scales}
\\
\\
\Delta_{\rm T}
=\displaystyle\frac{4}{27}\,
\left(\epsilon_{1}\,\epsilon_{2}^2
+\epsilon_{1}\,\epsilon_{2}\,\epsilon_{3} \right)
+\mathcal{O}(\epsilon_{n}^{4})
&
\quad
\textrm{super-horizon scales}
\ ,
\end{array}
\right.
\label{rapp_Q/omega2}
\ee
\end{widetext}
and it is thus obvious that $\Delta$ becomes very small for any
inflationary potential satisfying the slow-roll
conditions~\cite{wang_mukh} in both limits of interest
(but, remarkably, not at the turning point).
As a consequence, the WKB approximation is now also valid on
super-horizon scales.
\par
One might naively expect that the transformation~(\ref{transf})
will improve the WKB approach to all orders (in the formal
expansion in $\delta$), thus yielding increasingly
better results.
However, since $Q$ diverges at the turning point $x=x_*$,
whether the function $\mu_{\rm WKB}$ can still be regarded
as a solution to Eq.~(\ref{eq_wkb_exp}) for $x\sim x_*$ in
the adiabatic limit $\delta\to 0$ becomes a subtle issue.
It is precisely for this reason that one usually considers
particular solutions around the turning point (for example,
combinations of Airy functions for a linear turning point)
which must then be matched with the asymptotic
form~(\ref{wkb_sol}) on both sides of the turning point.
The problem with this standard procedure is that there is no
general rule to determine the actual matching points
(which cannot be $x_*$ where $Q$ diverges) in such a way that
the overall error be small.
As a consequence, the standard adiabatic expansion for
$\delta\ll 1$ may not lead to a reliable approximation
around the turning point, which would then result in large
errors to higher orders and a questionable further
extension of all the expressions to the interesting case
$\delta=1$.
\par
It was in fact shown in Ref.~\cite{hunt} that a simple
extension to higher orders of the above procedure~\cite{bender}
actually seems to reduce the accuracy with respect to the
leading order results of~Ref.~\cite{martin_schwarz}.
We shall see in the next Section that one can overcome this
problem by introducing mode functions which remain good
approximations also around the turning point.
\section{Improved WKB approximation}
\label{schiffmethod}
We shall now describe a method to improve the WKB approximation
to all orders by following the scheme outlined in Ref.~\cite{langer}
(see also Ref.~\cite{schiff}).
The basic idea is to use approximate expressions which are valid
for all values of the coordinate $x$ and then expand around them.
Two different such expansions we shall consider, namely
one based on the Green's function technique and the other
on the usual adiabatic expansion~\cite{langer}.
\par
Let us first illustrate the main idea of Ref.~\cite{langer} 
for the particular case $\omega^{2}=C\,\left(x-x_*\right)^{n}$,
with $C$ a positive costant.
All the solutions to Eq.~(\ref{new_eq}) for $x>x_*$ can be written
as linear combinations of the two functions
\be
u_{\pm}(x)=\sqrt{\frac{\xi(x)}{\omega(x)}}\,
J_{\pm m}\left[\xi(x)\right]
\ ,
\label{particular_sol}
\ee
where
\be
\xi(x):=\int_{x_*}^{x}\omega(y)\,\d y
\ ,
\quad
\quad
m=\frac{1}{n+2}
\ ,
\label{xi_and_m}
\ee
and $J_\nu$ are Bessel functions~\cite{abram}.
Moreover, for a general frequency $\omega$,
the expressions in Eq.~(\ref{particular_sol})
satisfy
\be
\left[\frac{{\d}^2}{{\d} x^2}+\omega^2(x)-\sigma(x)\right]
\,u_{\pm}=0
\ ,
\label{eq_with_eta}
\ee
where the quantity (primes from here on will denote
derivatives with respect to the argument of the given
function)
\be
\sigma(x)&=&\frac{3}{4}\,\frac{(\omega')^2}{\omega^2}
-\frac{\omega''}{2\,\omega}
+\left(m^{2}-\frac14\right)\,\frac{\omega^{2}}{\xi^{2}}
\nonumber
\\
&=&
Q_\omega+\left(m^{2}-\frac14\right)\,\frac{\omega^{2}}{\xi^{2}}
\ ,
\label{eta}
\ee
contains the term $Q$ defined in Eq.~(\ref{Q}),
whose divergent behavior at the turning point $x=x_*$ we
identified as the possible cause of failure of the standard WKB
approach.
For a general (finite) frequency, which can be expanded in powers
of $x-x_*$~\footnote{In the following we shall consider the
general case in Eq.~(\ref{omega_expand}), although one might restrict
to $n=1$, since our frequencies exhibit a linear turning point.},
\be
\omega^{2}(x)=C\,\left(x-x_*\right)^{n}
\left[1+\sum_{q\ge 1}\,c_q\,(x-x_*)^q\right]
\ ,
\label{omega_expand}
\ee
one finds that the extra term in Eq.~(\ref{eta}) precisely
``removes'' the divergence in $Q$ at the turning point.
In fact, the residue
\be
\lim_{x\to x_*}\sigma(x)=\frac{3\,(n+5)\,c_1^{2}}{2\,(n+4)\,(n+6)}
-\frac{3\,c_2}{n+6}
\ ,
\label{limit_eta_at_TP}
\ee
is finite~\cite{langer,schiff}.
The finiteness of $\sigma$ at the turning point is crucial in
order to extend the WKB method to higher orders
(this point was missed in Ref.~\cite{hunt}).
It is also easy to show that, for the cases of interest,
$|\sigma/\omega^2|\simeq\Delta$ as given in
Eq.~(\ref{rapp_Q/omega2}) for $x\to\pm\infty$, so that
the new approximate solutions remain very accurate
for large $|x|$.
For slowly varying frequencies, with $|\sigma(x)|$ small
everywhere, the functions~(\ref{particular_sol}) are thus
expected to be good approximations to the solutions of
Eq.~(\ref{new_eq}) for the whole range of $x$, including
the turning point.
\par
We can now introduce both the adiabatic expansion of the previous
Section and a new formal expansion of the mode functions by
replacing Eq.~(\ref{new_eq}) by
\be
\left[\delta\,\frac{{\d}^2}{{\d} x^2}+
\omega^2(x)-\delta\,\sigma(x)\right]\,\chi
=-\delta\,\varepsilon\,\sigma(x)\,\chi
\ ,
\label{new_eq_rewrite}
\ee
where $\delta$ and $\varepsilon$ are ``small'' positive
parameters.
We shall refer to expressions proportional to $\delta^n$ as
the $n$-th adiabatic order and to those proportional to
$\varepsilon^n$ as the $n$-th perturbative order.
\par
It is convenient to consider the solutions to
Eq.~(\ref{new_eq_rewrite}) to the left and right of the
turning point $x_*$ separately.
We shall call:
\begin{subequations}
\begin{description}
\item[region~I)]
where $\omega^{2}>0$ (on the right of $x_*$), with
\be
\omega_{\rm I}(x):=\sqrt{\omega^{2}(x)}
\ ,
\quad
\xi_{\rm I}(x):=\int_{x_*}^{x}\omega_{\rm I}(y)\,\d y
\ ,
\label{xi_I}
\ee
and the solutions to the corresponding homogeneous equation
(\ref{eq_with_eta}) are given in Eq.~(\ref{particular_sol});
\item[region~II)]
where $\omega^{2}<0$ (on the left of
$x_*$), with
\be
\omega_{\rm II}(x):=\sqrt{-\omega^{2}(x)}
\ ,
\quad
\xi_{\rm II}(x):=\int_{x}^{x_{*}}\omega_{\rm II}(y)\,\d y
\ ,
\label{xi_II}
\ee
and the solutions to the corresponding homogeneous equation
(\ref{eq_with_eta}) are obtained from those in
Eq.~(\ref{particular_sol}) by replacing the $J_\nu$ with the $I_\nu$~\cite{abram}.
\end{description}
\end{subequations}
Corresponding expressions are obtained for all relevant
quantities, for example $\sigma_{\rm I}$ and $\sigma_{\rm II}$,
and we shall omit the indices I and II whenever it is not
ambiguous.
\par
Although it is possible to expand in both parameters,
so as to obtain terms of order $\delta^p\,\varepsilon^q$
(with $p$ and $q$ positive integers),
in the following we shall just consider each expansion separately,
that is we shall set $\delta=1$ in the perturbative expansion
(see~Section~\ref{pert}) and $\varepsilon=1$ in the adiabatic
expansion (see~Section~\ref{adiab}).
%
%
%
\section{Perturbative expansion}
\label{pert}
For $\delta=1$, the limit $\varepsilon\to 0$ is exactly solvable
and yields the solutions (\ref{particular_sol}), whereas the case
of interest (\ref{new_eq}) is recovered in the limit
$\varepsilon\to 1$, which must always be taken at the end of
the computation.
The solutions to Eq.~(\ref{new_eq_rewrite}) can be further
cast in integral form as
\be
\chi(x)=u(x)-\varepsilon\,\int G(x,y)\,\sigma(y)\,\chi(y)\,\d y
\ ,
\label{complete_sol}
\ee
where $u(x)=A_+\,u_{+}(x)+A_-\,u_{-}(x)$ is a linear combination
of solutions (\ref{particular_sol}) to the corresponding homogeneous
equation~(\ref{eq_with_eta}), and $G(x,y)$ is the Green's function
determined by
\be
\left[\frac{{\d}^2}{{\d} x^2}+\omega^2(x)-\sigma(x)\right]
\,G(x,y)
=\varepsilon\,\delta(x-y)
\ .
\label{green_eq}
\ee
The solutions of the homogeneous equation~(\ref{eq_with_eta})
in the two regions are then given by
\be
u_{\rm I}(x)&\!\!\!=\!\!\!&
\displaystyle\sqrt{\frac{\xi_{\rm I}(x)}{\omega_{\rm I}(x)}}\,
\left\{
A_{+}\,J_{+m}\left[\xi_{\rm I}(x)\right]
+A_{-}\,J_{-m}\left[\xi_{\rm I}(x)\right]
\right\}
\nonumber
\\
&
\!\!\!:=\!\!\!
&
A_+\,u_{{\rm I}+}(x)+A_-\,u_{{\rm I}-}(x)
\nonumber
\\
\label{sol_in}
\\
u_{\rm II}(x)&\!\!\!=\!\!\!&
\displaystyle\sqrt{\frac{\xi_{\rm II}(x)}{\omega_{\rm II}(x)}}\,
\left\{
B_{+}\,I_{+m}\left[\xi_{\rm II}(x)\right]
+B_{-}\,I_{-m}\left[\xi_{\rm II}(x)\right]
\right\}
\nonumber
\\
&\!\!\!:=\!\!\!&
B_+\,u_{{\rm II}+}(x)+B_-\,u_{{\rm II}-}(x)
\ ,
\nonumber
\ee
and the Green's functions can be written as
\be
G_{\rm I}(x,y)&=&
-\displaystyle\frac{\pi}{\sqrt{3}}\,
\theta(x-y)\,u_{{\rm I}-}(x)\,u_{{\rm I}+}(y)
\nonumber
\\
&&
-\displaystyle\frac{\pi}{\sqrt{3}}\,
\theta(y-x)\,u_{{\rm I}+}(x)\,u_{{\rm I}-}(y)
\nonumber
\\
\label{green_funct}
\\
G_{\rm II}(x,y)&=&
\displaystyle\frac{i\,\pi}{\sqrt{3}}\,
\theta(x-y)\,u_{{\rm II}-}(x)\,u_{{\rm II}+}(y)
\nonumber
\\
&&
+\,\displaystyle\frac{i\,\pi}{\sqrt{3}}\,
\theta(y-x)\,u_{{\rm II}+}(x)\,u_{{\rm II}-}(y)
\ .
\nonumber
\ee
\par
With the ansatz
\be
\!\!\!\!\!\!\!\!\!\!
\chi_{\rm I}(x)&\!\!=\!\!&
A_{+}\,a_{+}(x)\,u_{{\rm I}+}(x)
+A_{-}\,a_{-}(x)\,u_{{\rm I}-}(x)
\nonumber
\\
\label{ansatz}
\\
\!\!\!\!\!\!\!\!\!\!
\chi_{\rm II}(x)&\!\!=\!\!&
B_{+}\,b_{+}(x)\,u_{{\rm II}+}(x)
+B_{-}\,b_{-}(x)\,u_{{\rm II}-}(x)
\ ,
\nonumber
\ee
from Eqs.~(\ref{sol_in}) and~(\ref{green_funct}),
the problem is now reduced to the determination
of the $x$-dependent coefficients
\be
&&
\!\!\!\!\!\!\!\!\!\!
a_{+}(x):=\left[1+\varepsilon\,J^+_{+-}(x,x_i)\right]
+\varepsilon\,\frac{A_{-}}{A_+}\,J^-_{--}(x,x_i)
\nonumber
\\
\nonumber
\\
&&
\!\!\!\!\!\!\!\!\!\!
a_{-}(x):=
\left[1+\varepsilon\,J^-_{+-}(x_*,x)\right]
+\varepsilon\,\frac{A_{+}}{A_-}\,J^+_{++}(x_*,x)
\nonumber
\\
\label{Abar_Bbar}
\\
&&
\!\!\!\!\!\!\!\!\!\!
b_{+}(x):=
\left[1-i\,\varepsilon\,I^+_{+-}(x,x_*)\right]
-i\,\varepsilon\,\frac{B_{-}}{B_+}\,I^-_{--}(x,x_*)
\nonumber
\\
\nonumber
\\
&&
\!\!\!\!\!\!\!\!\!\!
b_{-}(x):=
\left[1-i\,\varepsilon\,I^-_{+-}(x_f,x)\right]
-i\,\varepsilon\,\frac{B_{+}}{B_-}\,I^+_{++}(x_f,x)
\ ,
\nonumber
\ee
in which, for the sake of brevity (and clarity),
we have introduced a shorthand notation for the following
integrals
\be
&&\!\!\!\!\!\!\!\!\!\!
J^{w}_{s\bar s}(x_1,x_2):=
\displaystyle\frac{\pi}{\sqrt{3}}\,
\int_{x_1}^{x_2}
\sigma_{\rm I}(y)\,a_w(y)\,
u_{{\rm I}s}(y)\,u_{{\rm I}\bar s}(y)\,\d y
\nonumber
\\
&&
\!\!\!\!\!\!\!\!\!\!
=
\displaystyle\frac{\pi}{\sqrt{3}}\,\int_{x_1}^{x_2}
\!\!\!\!
\sigma_{\rm I}(y)
\,\frac{\xi_{\rm I}(y)}{\omega_{\rm I}(y)}\,a_{w}(y)\,
\,J_{sm}\left[\xi_{\rm I}(y)\right]\,J_{\bar sm}\left[\xi_{\rm I}(y)\right]
\d y
\nonumber
\\
\phantom{A}
\\
&&\!\!\!\!\!\!\!\!\!\!
I^{w}_{s\bar s}(x_1,x_2):=
\displaystyle\frac{\pi}{\sqrt{3}}\,
\int_{x_1}^{x_2}
\sigma_{\rm II}(y)\,b_w(y)\,
u_{{\rm II}s}(y)\,u_{{\rm II}\bar s}(y)
\,\d y
\nonumber
\\
&&
\!\!\!\!\!\!\!\!\!\!\!\!
=
\displaystyle\frac{\pi}{\sqrt{3}}\,\int_{x_1}^{x_2}
\!\!\!\!
\sigma_{\rm II}(y)
\,\frac{\xi_{\rm II}(y)}{\omega_{\rm II}(y)}\,b_{w}(y)\,
\,I_{sm}\left[\xi_{\rm II}(y)\right]\,
I_{\bar sm}\left[\xi_{\rm II}(y)\right]
\d y
\, ,
\nonumber
\ee
where the symbols $w$, $s$ and $\bar s$ are $+$ or $-$.
\par
Before tackling this problem, we shall work out relations
which allow us to determine the constant coefficients $A_\pm$
and $B_\pm$ uniquely.
\subsection{Initial and matching conditions}
We first use the initial conditions~(\ref{init_cond_on_mu})
in order to fix the constant coefficients $A_\pm$.
\par
For the function $\chi$ such conditions become
\be
\lim_{x\rightarrow x_i}\chi(x)
&=&
d\,\sqrt{\frac{1-\epsilon_{1}(x_i)}{2k}}\,
{\rm e}^{-i\,k\,\eta_i-x_i/2}
\nonumber
\\ &=& 
d\,\sqrt{\frac{1-\epsilon_{1}(x_i)}{2k}}\,
{\rm e}^{i\,\xi_{\rm I}(x_i)-x_i/2}
\ ,
\label{init_cond_on_chi}
\ee
where, from the definitions of $\eta$ and $x$,
$k\,\eta_i\simeq -i\,\xi_{\rm I}(x_i)$.
We then recall the asymptotic form of the Bessel functions $J_\nu$
for $x\to+\infty$~\cite{abram,schiff}~\footnote{Since the initial
conditions must be imposed at $x$ in deep region~I, 
we can consider $x_i\sim+\infty$.},
\be
J_{\pm m}\left[\xi_{\rm I}(x)\right]
\sim
\sqrt{\frac{2}{\pi\,\xi_{\rm I}(x)}}
\,\cos\left(\xi_{\rm I}(x)\mp\frac{\pi}{2}\,m-\frac{\pi}{4}\right)
\ ,
\label{J_asymp}
\ee
from which we obtain the asymptotic expression of the mode
function in region~I,
\begin{widetext}
\be
\chi_{\rm I}(x_i)&\!\!\sim\!\!&
\sqrt{\frac{1-\epsilon_{1}(x_i)}{2\,\pi}}\,{\rm e}^{-{x_i}/{2}}\,
\left\{{\rm e}^{+i\,\xi_{\rm I}(x_i)}
\,\left[{\rm e}^{-i\,\left(\frac{1}{4}+\frac{m}{2}\right)\,\pi}\,A_{+}
+{\rm e}^{-i\,\left(\frac{1}{4}-\frac{m}{2}\right)\,\pi}
\left(A_{-}+\varepsilon\,A_{-}\,J_{+-}^{-}(x_*,x_i)
+\varepsilon\,A_{+}\,J_{++}^{+}(x_*,x_i)\right)
\right]\right.
\nonumber
\\
&&
\ \ \ \ \ \ \ \ \ \ \ \ \
\left.+{\rm e}^{-i\,\xi_{\rm I}(x_i)}\,
\left[{\rm e}^{+i\,\left(\frac{1}{4}+\frac{m}{2}\right)\,\pi}\,A_+
+{\rm e}^{+i\,\left(\frac{1}{4}-\frac{m}{2}\right)\,\pi}\,
\left(A_{-}+\varepsilon\,A_{-}\,J_{+-}^{-}(x_*,x_i)
+\varepsilon\,A_{+}\,J_{++}^{+}(x_*,x_i)\right)
\right]
\right\}
\ .
\label{init_sol in_I}
\ee
\end{widetext}
The initial conditions~(\ref{init_cond_on_chi}) therefore yield
\begin{subequations}
\be
&&
\!\!\!\!\!\!
A_{+}
=d\,\sqrt{\frac{\pi}{k}}\,
\frac{{\rm e}^{+i\,\left(\frac{1}{4}+\frac{m}{2}\right)\,\pi}}
{1-{\rm e}^{2\,i\,m\,\pi}} 
\label{A+}
\\
&&
\!\!\!\!\!\!
\phantom{A}
\nonumber
\\
&&
\!\!\!\!\!\!
A_{-}
=-A_{+}\,\displaystyle\frac{{\rm e}^{+i\,m\,\pi}
+\varepsilon\,J_{++}^{+}(x_*,x_i)}
{1+\varepsilon\,J_{+-}^{-}(x_*,x_i)}
\ .
\label{A-}
\ee
\end{subequations}
Note that $A_+$ does not depend on $\varepsilon$, therefore the result
(\ref{A+}) holds to all orders.
\par
We next impose continuity at the turning point in order to determine
the coefficients $B_\pm$.
For this we shall need the (asymptotic) expressions of Bessel functions
for $x\to x_*$~\cite{abram,schiff}~\footnote{The Bessel functions are
actually regular around $x=x_*$, therefore the following expressions
are just the leading terms in the Taylor expansion of $J_\nu$ and $I_\nu$.}.
In particular,
\be
\begin{array}{ll}
J_{\pm m}\left[\xi_{\rm I}(x)\right]
\simeq
\displaystyle
\left[\frac{\xi_{\rm I}(x)}{2}\right]^{\pm m}\,
\frac{1}{\Gamma\left(1\pm m\right)}
&
{\rm for}\
x\to x_*^+
\\
\\
I_{\pm m}\left[\xi_{\rm II}(x)\right]
\simeq
\displaystyle
\left[\frac{\xi_{\rm II}(x)}{2}\right]^{\pm m}\,
\frac{1}{\Gamma\left(1\pm m\right)}
&
{\rm for}\
x\to x_*^-
\ ,
\end{array}
\label{JI_taylor}
\ee
where $\Gamma$ is Euler's gamma function.
Near the turning point, we just keep the leading term in the
expansion~(\ref{omega_expand}), so that
\be
&&
\omega(x)\simeq\displaystyle
\sqrt{C}\,\left|x-x_*\right|^{\frac{1-2\,m}{2\,m}}
\nonumber
\\
\label{DX_TP}
\\
&&
\xi(x)\simeq
2\,m\,{\sqrt{C}}\,\left|x-x_*\right|^{\frac{1}{2\,m}}
\ .
\nonumber
\ee
We thus obtain the following forms of the mode functions near
the turning point,
\begin{subequations}
\be
\!\!\!\!
\chi_{\rm I}(x\simeq x_*)\!\!&\simeq&\!\!
A_{+}\,a_+(x_*)\,
\frac{\sqrt{2\,m}\,\left(m\,\sqrt{C}\right)^{m}}
{\Gamma\left(1+m\right)}\,
\left(x-x_*\right)
\nonumber
\\
&&
\!\!\!\!
+A_{-}\,a_-(x_*)\,
\frac{\sqrt{2\,m}\,\left(m\,\sqrt{C}\right)^{-m}}
{\Gamma\left(1-m\right)}
\label{sol_IX_at_TP}
\\
\phantom{A}
\nonumber
\\
\!\!\!\!\!\!\!\!
\chi_{\rm II}(x\simeq x_*)\!\!&\simeq&\!\!
B_{+}\,b_+(x_*)\,
\frac{\sqrt{2\,m}\,\left(m\,\sqrt{C}\right)^{m}}
{\Gamma\left(1+m\right)}\,
\left(x_*-x\right)
\nonumber
\\
&&
\!\!\!\!
+B_{-}\,b_-(x_*)\,
\frac{\sqrt{2\,m}\,\left(m\,\sqrt{C}\right)^{-m}}
{\Gamma\left(1-m\right)}
\ .
\label{sol_IIX_at_TP}
\ee
\end{subequations}
Continuity across $x_*$ then implies that
\be
B_{+}\,b_+(x_*)=
-A_{+}\,a_+(x_*)
\nonumber
\\
\phantom{A}
\label{connection_at_TP}
\\
B_{-}\,b_-(x_*)=
A_{-}\,a_-(x_*)
\ .
\nonumber
\ee
From the definitions~(\ref{Abar_Bbar}) at $x=x_*$,
Eq.~(\ref{A-}) and the relations~(\ref{connection_at_TP})
we therefore obtain
\be
\frac{B_{+}}{A_+}&\!\!=\!\!&
-1
+\varepsilon\,
\frac{{\rm e}^{+i\,m\,\pi}
+\varepsilon\,J_{++}^{+}(x_*,x_i)}
{1+\varepsilon\,J_{+-}^{-}(x_*,x_i)}\,
J_{--}^{-}(x_*,x_i)
\nonumber
\\
&&
-\varepsilon\,J_{+-}^{+}(x_*,x_i)
\nonumber
\\
\label{B+-}
\\
\frac{B_{-}}{A_+}&\!\!=\!\!&
-\frac{{\rm e}^{+i\,m\,\pi}
+\varepsilon\,J_{++}^{+}(x_*,x_i)}
{1+\varepsilon\,J_{+-}^{-}(x_*,x_i)}
\nonumber
\\
&&
\ \
\times
\frac{1-i\,\varepsilon^2\,J_{--}^{-}(x_*,x_i)\,
I_{++}^{+}(x_f,x_*)}
{1-i\,\varepsilon\,I_{+-}^{-}(x_f,x_*)}
\nonumber
\\
&&
-i\,\varepsilon\,
\displaystyle\frac{1+\varepsilon\,J_{+-}^{+}(x_*,x_i)}
{1-i\,\varepsilon\,I_{+-}^{-}(x_f,x_*)}\,I_{++}^{+}(x_f,x_*)
\ .
\nonumber
\ee
\par
It is apparent from the above derivation that, since the matching
between approximate solutions in the two regions is always
performed at $x=x_*$, no ambiguity hinders the evaluation of
the coefficients $B_\pm$ and the accuracy of the approximate
solution is therefore expected to increase with increasing order.
\subsection{Recursive (perturbative) relations}
We can finally obtain expressions for the coefficients
${a}_{\pm}(x)$ and ${b}_{\pm}(x)$ by solving the integral
relations (\ref{Abar_Bbar}).
We remark that the r.h.s.'s of such equations contain
$a_\pm(x)$ and $b_\pm(x)$, which can therefore be determined
recursively by expanding them (as well as any function of them)
in $\varepsilon$, that is
\be
&&
a_\pm(x)=1+\displaystyle\sum_{q\geq1} \varepsilon^q\,a_{\pm}^{(q)}
\nonumber
\\
\\
&&
b_\pm(x)=1+\displaystyle\sum_{q\geq1} \varepsilon^q\,b_{\pm}^{(q)}
\ .
\nonumber
\ee
It will also be useful to define the following integrals
\be
&&
\!\!\!\!\!\!\!\!\!\!
J^{w(q)}_{s\bar s}(x_1,x_2):=
\displaystyle\frac{\pi}{\sqrt{3}}\,
\int_{x_1}^{x_2}\!\!\!
\sigma_{\rm I}(y)\,a_w^{(q)}(y)\,
u_{Is}(y)\,u_{I\bar s}(y)\,\d y
\nonumber
\\
\\
&&
\!\!\!\!\!\!\!\!\!\!
I^{w(q)}_{s\bar s}(x_1,x_2):=
\displaystyle\frac{\pi}{\sqrt{3}}\,
\int_{x_1}^{x_2}\!\!\!\!
\sigma_{\rm II}(y)\,b_w^{(q)}(y)\,
u_{IIs}(y)\,u_{II\bar s}(y)\,\d y
\,,
\nonumber
\ee
where again $w$, $s$ and $\bar s=\pm$, and $q$ is a
non-negative integer.
%
%
%
%
%
\subsubsection{Leading order}
\label{leading_results}
We begin by considering the leading~($\varepsilon^0$) order
\be
a_\pm^{(0)}(x)=b_\pm^{(0)}(x)=1
\ .
\label{0-th}
\ee
The corresponding solutions are simply given by
\begin{subequations}
\be
&&\!\!\!\!\!\!\!\!
\chi_{\rm I}(x)
\simeq
u_{\rm I}(x)
\label{sol_in_I_leading}
\\
&&
\!\!\!\!\!\!\!\!
=A_+\,\sqrt{\frac{\xi_{\rm I}(x)}{\omega_{\rm I}(x)}}\,
\left\{
J_{+m}\left[\xi_{\rm I}(x)\right]
-\,{\rm e}^{+i\,m\,\pi}\,J_{-m}\left[\xi_{\rm I}(x)\right]
\right\}
\nonumber
\\
\nonumber
\\
&&\!\!\!\!\!\!\!\!
\chi_{\rm II}(x)\simeq
u_{\rm II}(x)
\label{sol_in_II_leading}
\\
&&
\!\!\!\!\!\!\!\!
=-A_+\,\sqrt{\frac{\xi_{\rm II}(x)}{\omega_{\rm II}(x)}}\,
\left\{
\,I_{+m}\left[\xi_{\rm II}(x)\right]
+{\rm e}^{+i\,m\,\pi}\,I_{-m}\left[\xi_{\rm II}(x)\right]
\right\}
\ ,
\nonumber
\ee
\end{subequations}
%
where we used the zero~order forms of
Eqs.~(\ref{A-}) and~(\ref{B+-}),
\be
\begin{array}{l}
A_{-}=B_{-}=-A_{+}\,{\rm e}^{+i\,m\,\pi}
\\
\\
B_{+}=-A_{+}
\ ,
\end{array}
\label{relation_AB_leading}
\ee
and $A_+$ is given in Eq.~(\ref{A+}).
\par
We know that these expressions are also good solutions
of Eq.~(\ref{new_eq}) near the turning point and,
deep in region~II, we can use the asymptotic form of
$I_\nu$ for $x\sim x_f\to -\infty$~\cite{schiff,abram},
\be
I_{\pm m}\left[\xi_{\rm II}(x)\right]
&\!\!\sim\!\!&
\frac{1}{\sqrt{2\,\pi\,\xi_{\rm II}(x)}}
\nonumber
\\
&&
\times
\,\left[{\rm e}^{\xi_{\rm II}(x)}
+{\rm e}^{-\xi_{\rm II}(x)-i\,\left(\frac12\pm m\right)\,\pi}\right]
\ .
\label{I_asymp}
\ee
On neglecting the non-leading mode ${\rm e}^{-\xi_{\rm II}(x)}$,
we then obtain
\be
\chi_{\rm II}(x_f)\sim
-A_{+}
\,\frac{\left(1+{\rm e}^{+i\,m\,\pi}\right)}
{\sqrt{2\,\pi\,\omega_{\rm II}(x_f)}}
\,{\rm e}^{\xi_{\rm II}(x_f)}
\ .
\label{sol_deep_in II}
\ee
If we now use Eqs.~(\ref{spectra_def}),~(\ref{A+})
and~(\ref{sol_deep_in II}), and the values of $d$,
we recover the results of Ref.~\cite{martin_schwarz}
for all quantities of interest,
i.e.~power spectra, spectral indices and
$\alpha$-runnings
\begin{subequations}
\be
&&
\!\!\!\!\!\!\!\!\!\!\!\!\!\!\!\!\!
{\cal P}_{\zeta}
\simeq
\frac{H^2}{\pi\,\epsilon_1\,m_{\rm Pl}^2}
\left(\frac{k}{a\,H}\right)^3
\frac{{\rm e}^{2\,\xi_{{\rm II,S}}(k,\eta)}}
{\left[1-\epsilon_1(\eta)\right]\,\omega_{\rm II,S}(k,\eta)}
\nonumber
\\
&&
\!\!\!\!\!\!\!\!\!
:={\cal P}_{\zeta}^{(0)}
\label{spectra_S_M&S}
\\
\nonumber
\\
&&
\!\!\!\!\!\!\!\!\!\!\!\!\!\!\!\!\!
n_{\rm S}-1\simeq
3+2\,\left.\frac{\d\,\xi_{{\rm II,S}}}{\d\,\ln k}
\right|_{k=k_*}
\label{indices_S_M&S}
\\
\nonumber
\\
&&
\!\!\!\!\!\!\!\!\!\!\!\!\!\!\!\!\!
\alpha_{\rm S}\simeq
2\,\left.
\frac{\d^2\,\xi_{{\rm II,S}}}{\left(\d\,\ln k\right)^2}
\right|_{k=k_*}
\ ,
\label{runnings_S_M&S}
\ee
and
\be
&&
\!\!\!\!\!\!\!\!\!\!\!\!\!\!\!\!\!
{\cal P}_{h}\simeq
\frac{16\,H^2}{\pi\,m_{\rm Pl}^2}
\,\left(\frac{k}{a\,H}\right)^3\,
\frac{{\rm e}^{2\,\xi_{{\rm II,T}}(k,\eta)}}
{\left[1-\epsilon_1(\eta)\right]\,\omega_{\rm II,T}(k,\eta)}
\nonumber
\\
&&
\!\!\!\!\!\!\!\!\!
:={\cal P}_{h}^{(0)}
\label{spectra_T_M&S}
\\
\nonumber
\\
&&
\!\!\!\!\!\!\!\!\!\!\!\!\!\!\!\!\!
n_{\rm T}\simeq
3+2\,\left.\frac{\d\,\xi_{{\rm II,T}}}{\d\,\ln k}
\right|_{k=k_*}
\label{indices_T_M&S}
\\
\nonumber
\\
&&
\!\!\!\!\!\!\!\!\!\!\!\!\!\!\!\!\!
\alpha_{\rm T}\simeq
2\,\left.\frac{\d^2\,\xi_{{\rm II,T}}}{\left(\d\,\ln k\right)^2}
\right|_{k=k_*}
\ ,
\label{runnings_T_M&S}
\ee
\end{subequations}
%
where we have transformed back to the original
variables $k$ and $\eta$, and all quantities are
evaluated in the super-horizon limit (i.e.~for
$k\ll a\,H$).
\subsubsection{Next-to-leading order}
\label{next-to-leading_results}
We now insert the leading order expressions~(\ref{0-th})
into the integral relations~(\ref{Abar_Bbar}) to compute the
next-to-leading order expressions
%
\be
&&
{a}_{+}^{(1)}(x)=
J_{+-}^{(0)}(x,x_i)
-{\rm e}^{+i\,m\,\pi}\,J^{(0)}_{--}(x,x_i)
\nonumber 
\\
\nonumber
\\
&&
{a}_{-}^{(1)}(x)=
J^{(0)}_{+-}(x_*,x)
-{\rm e}^{-i\,m\,\pi}\,J^{(0)}_{++}(x_*,x)
\nonumber 
\\
\label{ab+-1}
\\
&&
{b}_{+}^{(1)}(x)=
-i\,I^{(0)}_{+-}(x,x_*)
-i\,{\rm e}^{+i\,m\,\pi}\,I^{(0)}_{--}(x,x_*)
\nonumber 
\\
\nonumber
\\
&&
{b}_{-}^{(1)}(x)=
-i\,I^{(0)}_{+-}(x_f,x)
-i\,{\rm e}^{-i\,m\,\pi}\,I^{(0)}_{++}(x_f,x)
\ ,
\nonumber 
\ee
%
where we used the simplified notation
$J^{+(0)}_{s\bar s}=J^{-(0)}_{s\bar s}=:J^{(0)}_{s\bar s}$
and $I^{+(0)}_{s\bar s}=I^{-(0)}_{s\bar s}=:I^{(0)}_{s\bar s}$,
and also expanded the coefficients $A_-$ and $B_\pm$ to zero
order in $\varepsilon$.
\par
From Eq.~(\ref{I_asymp}), again neglecting the non-leading mode
${\rm e}^{-\xi_{\rm II}(x)}$, we obtain the mode function
at $x=x_f$ to first order in $\varepsilon$,
\be
\!\!\!\!\!
\chi_{\rm II}(x_f)&\!\!\sim\!\!&
-\frac{A_+\,{\rm e}^{\xi_{\rm II}(x_f)}}
{\sqrt{2\,\pi\,\omega_{\rm II}(x_f)}}\,
\left\{1+{\rm e}^{+i\,m\,\pi}
\phantom{\frac{A}{B}}
\right.
\nonumber
\\
&&
\ \ \
+i\,\varepsilon\,
\left[I_{++}^{(0)}(x_f,x_*)-I_{+-}^{(0)}(x_f,x_*)\right]
\nonumber
\\
&&
\ \ \
-i\,\varepsilon\,{\rm e}^{+i\,m\,\pi}\,
\left[I_{--}^{(0)}(x_f,x_*)-I_{+-}^{(0)}(x_f,x_*)\right]
\nonumber
\\
&&
\ \ \
-\varepsilon\,{\rm e}^{+i\,m\,\pi}\,
\left[J_{+-}^{(0)}(x_*,x_i)+J_{--}^{(0)}(x_*,x_i)\right]
\nonumber
\\
&&
\left.
\phantom{\frac{A}{B}}
+\varepsilon\left[J_{+-}^{(0)}(x_*,x_i)+J_{++}^{(0)}(x_*,x_i)\right]
\right\}
\ ,
\label{sol_deep_in II_next_leading}
\ee
which, using Eqs.~(\ref{spectra_def}), (\ref{A+}),
(\ref{sol_deep_in II_next_leading}) and the values of $d$,
yields all quantities of interest to first order in
$\varepsilon$.
For the spectra we obtain
\begin{subequations}
\be
&&
\!\!\!\!\!\!\!\!\!\!\!\!\!\!\!\!
{\cal P}_{\zeta}\simeq
{\cal P}_{\zeta}^{(0)}\,
\left[1+g_{(1) \, {\rm S}}^{\rm GREEN} (x_f)\right]
\nonumber
\\
\label{spectra_next-to-leading}
\\
&&
\!\!\!\!\!\!\!\!\!\!\!\!\!\!\!\!
{\cal P}_{h}\simeq
{\cal P}_{h}^{(0)}\,
\left[1+g_{(1) \, {\rm T}}^{\rm GREEN} (x_f)\right]
\ ,
\nonumber
\ee
where the relative corrections to the leading order
expressions~(\ref{spectra_S_M&S}) and~(\ref{spectra_T_M&S})
are now given by
\be
&&
\!\!\!\!\!\!\!\!
g_{(1) \, {\rm S,T}}^{\rm GREEN} (x)=
\left\{
J_{++}^{(0)}(x_*,x_i)-J_{--}^{(0)}(x_*,x_i)
\right.
\label{g(1)}
\\
&&
\!\!\!\!\!\!\!\!
\left.
+\frac{1}{\sqrt{3}}
\left[I_{++}^{(0)}(x,x_*)
-2\,I_{+-}^{(0)}(x,x_*)
+I_{--}^{(0)}(x,x_*)\right]
\right\}_{\rm S,T}
\ ,
\nonumber
\ee
\end{subequations}
in which we set $\varepsilon=1$ and $n=1$ as required, and
write S and T to recall the use of the corresponding
frequencies.
The expressions for the spectral indices and their runnings
to this order can finally be derived from the
definitions~(\ref{n_def}) and~(\ref{alpha_def}).
\subsubsection{Higher orders}
\label{next-to-next-to-leading_results}
The above procedure can be extended to all orders.
The coefficients up to $a_\pm^{(q)}$ and $b_\pm^{(q)}$ must
be used to compute the integrals $J^{w(q)}_{s\bar s}$ and
$I^{w(q)}_{s\bar s}$ which determine $a_\pm^{(q+1)}$ and
$b_\pm^{(q+1)}$.
However, general expressions soon become very involved,
and we shall test the effectiveness of our method by
applying it to a case of interest in
Section~\ref{Power-law inflation}.
\section{Adiabatic expansion}
\label{adiab}
Let us now apply the usual adiabatic expansion with the
assumption that the leading order be given by the
functions~(\ref{particular_sol})
[rather than the more common expression~(\ref{wkb_sol})].
This leads one to replace
\be
\xi(x)\to 
\frac{\xi(x)}{\sqrt{\delta}}
:=\frac{1}{\sqrt{\delta}}\,\int^x
\omega(y)\,\d y
\ ,
\ee
and consider the forms~\cite{langer}.
\begin{subequations}
\be
U_\pm=F_\pm(x;\delta)\,u_\pm(x)+G_\pm(x;\delta)\,u'_\pm(x)
\ ,
\ee
where the $x$-dependent coefficients $F$ and $G$ must now
be determined.
The particular case of a linear turning
point~\cite{langer49}
(precisely the one which mostly concerns us here),
can then be fully analyzed as follows.
\subsection{Recursive (adiabatic) relations}
The $x$-dependent coefficients $F$ and $G$ can be
expanded in powers of $\delta$ as
\be
&&
F(x;\delta)=\sum_{j\ge 0} \delta^j\,\phi_{(j)}(x)
\nonumber
\\
\label{adiab_exp}
\\
&&
G(x;\delta)=\sum_{j\ge 0} \delta^j\,\gamma_{(j)}(x)
\nonumber
\ .
\ee
Upon substituting into Eq.~(\ref{new_eq_rewrite}) with
$\varepsilon=1$, one therefore obtains that the coefficients
$\phi_{(j)}$ and $\gamma_{(j)}$ are given recursively
by the formulae
\be
\phi_{(j)}(x)&\!\!=\!\!&-\frac{1}{2}
\int^{x}
\left[\gamma''_{(j)}(y)+\sigma(y)\,\gamma_{(j)}(y)\right]
\d y
\nonumber
\\
\label{delta_j}
\\
\gamma_{(j)}(x)&\!\!=\!\!&\frac{1}{2\,\omega(x)}
\int^x
\!\!
\left\{
\sigma(y)\,\left[2\,\gamma'_{(j-1)}(y)+\phi_{(j-1)}(y)\right]
\right.
\nonumber
\\
&&
\phantom{\frac{1}{2\,\omega(x)}\,\int\!\!\!}
\left.
+\sigma'(y)\,\gamma_{(j-1)}(y)+\phi''_{(j-1)}(y)\right\}
\frac{\d y}{\omega(y)}
\ ,
\nonumber
\ee
\end{subequations}
where it is understood that the integration must be performed
from $x_*$ to $x$ in region~I and from $x$ to $x_*$ in region~II.
Let us also remark that, for $\omega^2\sim (x-x_*)$ near the
turning point, the above expressions are finite~\cite{langer49},
whereas for more general cases one expects divergences as
with the more standard WKB approach.
\subsubsection{Leading order}
The lowest order solutions (\ref{particular_sol}) are correctly
recovered on setting 
\be
\phi_{(0)}=1
\quad
{\rm and}
\quad
\gamma_{(0)}=0
\ ,
\label{delta0}
\ee
in the limit $\delta=1$.
The relevant leading order perturbations are therefore obtained
on imposing the initial conditions~(\ref{init_cond_on_chi})
and matching conditions between region~I and region~II at the
turning point, which therefore yield for $\chi_{\rm I}$ and
$\chi_{\rm II}$ the same linear combinations $u_{\rm I}$
of Eq.~(\ref{sol_in_I_leading}) and $u_{\rm II}$ of
Eq.~(\ref{sol_in_II_leading}) previously obtained.
\subsubsection{Higher orders}
\label{adiab_1}
The condition~(\ref{delta0}) makes the formal expressions of
the coefficients $\phi_{(1)}$ and $\gamma_{(1)}$ particularly
simple,
\be
\phi_{(1)}(x)&\!\!=\!\!&-\frac{1}{2}\,\int^{x}
\left[\gamma''_{(1)}(y)+\sigma(y)\,\gamma_{(1)}(y)\right]\,
\d y
\nonumber
\\
\label{delta1}
\\
\gamma_{(1)}(x)&\!\!=\!\!&\frac{1}{2\,\omega(x)}\,\int^x
\frac{\sigma(y)}{\omega(y)}\,
\d y
\ ,
\nonumber
\ee
and the first order solutions are then given by linear combinations
of the two functions
\be
U_\pm(x)=\left[1+\delta\,\phi_{(1)}(x)\right]\,u_\pm(x)
+\delta\,\gamma_{(1)}(x)\,u'_\pm(x)
\ .
\ee
\par
It is now important to observe that the initial conditions~(\ref{init_cond_on_chi}) and matching conditions
at the turning point to first order in the adiabatic
parameter $\delta$ yield different results with respect
to the perturbative expansion in $\varepsilon$.
First of all, the coefficient $A_+$, which was left unaffected
by the expansion in $\varepsilon$ [see~Eq.~(\ref{A+})],
acquires a correction.
The relations between $A_-$, $B_\pm$ and $A_+$ instead
remain those given in Eq.~(\ref{relation_AB_leading})
to all orders, whereas in the perturbative
expansion they were modified [see Eqs.~(\ref{A-})
and~(\ref{B+-})].
To summarize, to first order in $\delta$,
one obtains
\be
A_+&\!\!\simeq\!\!&
d\,\sqrt{\frac{\pi}{k}}\,
\frac{{\rm e}^{i\,\left(\frac{1}{4}+\frac{m}{2}\right)\,\pi}}
{1-{\rm e}^{2\,i\,m\,\pi}}
\nonumber
\\
&&
\times
\left\{1-\delta\,\left[
\phi_{{\rm I}(1)}(x_i)
+\gamma_{{\rm I}(1)}(x_i)\,\left(
i\,\omega_{\rm I}(x_i)-\frac{1}{2}\right)
\right]\right\}
\nonumber
\\
\nonumber
\\
A_-&\!\!=\!\!&
-A_+\,{\rm e}^{+i\,m\,\pi}
\label{ABadia1}
\\
\nonumber
\\
B_\pm&\!\!=\!\!&
\mp A_\pm
\ ,
\nonumber
\ee
and the perturbation modes deep in Region~II 
(at $x\ll x_*$) are finally given by
%
\begin{widetext}
\be
\chi_{\rm II}(x)
&\simeq&
u_{\rm II}(x)\,
\left\{1+
\delta\,\left[
\phi_{{\rm II}(1)}(x)
-\gamma_{{\rm II}(1)}(x)\,\left(
\omega_{\rm II}(x)
+\frac{\omega_{\rm II}'(x)}{2\,\omega_{\rm II}(x)}\right)
-\phi_{{\rm I}(1)}(x_i)
+\gamma_{{\rm I}(1)}(x_i)\,\left(
\frac{1}{2}-i\,\omega_{\rm I}(x_i)\right)
\right]
\right\}
\ .
\label{chi_1}
\ee
\end{widetext}
We note that, in order to obtain the above result,
one must first take the asymptotic expansion of the
functions $u_\pm$ and then take the derivative,
since the reverse order would lead to larger errors.
The corrected power spectra are then given by
\begin{subequations}
\be
&&
{\cal P}_\zeta\simeq
{\cal P}^{(0)}_\zeta\,\left[1+g_{(1){\rm S}}^{\rm AD}(x_f)
\right]
\nonumber
\\
\\
&&
{\cal P}_h\simeq
{\cal P}^{(0)}_h\,\left[1+g_{(1){\rm T}}^{\rm AD} (x_f)\right]
\ ,
\nonumber
\ee
in which we set $\delta=1$ at the end,
and~\footnote{Since our method has some similarity with that
of Ref.~\cite{HHHJM} (see also Ref.~\cite{olver}),
it is worth pointing out that the first order corrections
shown here differ from those of Ref.~\cite{HHHJM} (at least)
in that they contain contributions from both Regions~I and~II.} 
\be
g_{(1){\rm S,T}}^{\rm AD}(x)&\!\!\!=\!\!\!&
2\left[
\phi_{{\rm II}(1)}(x)
-\gamma_{{\rm II}(1)}(x)
\left(\omega_{\rm II}(x)
+\frac{\omega_{\rm II}'(x)}{2\,\omega_{\rm II}(x)}\right)
\right.
\nonumber
\\
&&
\left.
\phantom{2\,[}
+\frac{\gamma_{{\rm I}(1)}(x_i)}{2}
-\phi_{{\rm I}(1)}(x_i)
\right]_{\rm S,T}
\ ,
\label{g_1}
\ee
\end{subequations}
where the indices S and T recall the use of the
corresponding frequencies.
As usual, the corrections for the spectral indices and
$\alpha$-runnings then follow from their
definitions~(\ref{n_def}) and~(\ref{alpha_def}).
\par
Given the complicated expressions for $\omega$, it is
usually impossible to carry out the double integration that
yields $\phi_{(1)}$ analytically, let alone higher order terms.
One must therefore rely on numerical computations.
In Section~\ref{Power-law inflation}, we shall, in the
context of power-law inflation, compare the results obtained
in next-to-leading order from the two expansions so far
described in full generality.
\section{Application to Power-law Inflation}
\label{Power-law inflation}
\begin{figure}[ht]
\centerline{\raisebox{2.5cm}{$\omega^2\ $}
\includegraphics[width=0.4\textwidth]{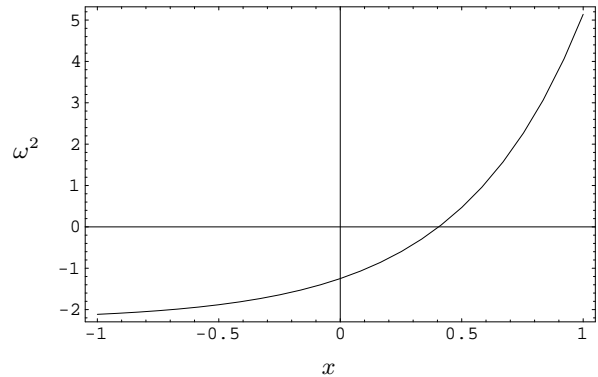}}
\centerline{$\ \ \ \ \ \ x$}
\caption{The frequency $\omega^2$ for $\beta=-2$.
The turning point is at $x_*=\ln(3/2)\simeq 0.4$}
\label{omega2_deSitt}
\end{figure}

\begin{figure}[h]
\centerline{\raisebox{2.5cm}{$\xi\ $}
\includegraphics[width=0.4\textwidth]{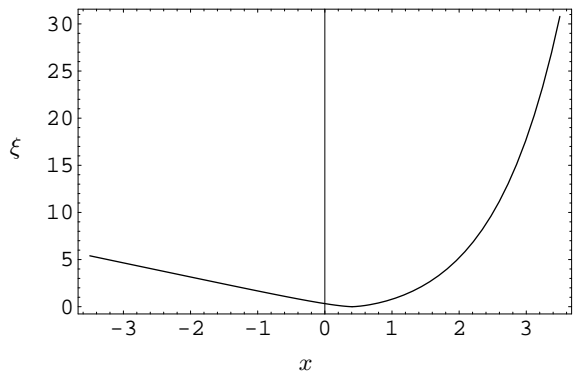}}
\centerline{$\ \ \ x$}
\caption{The function $\xi$ for $\beta=-2$.}
\label{xi_deSitt}
\end{figure}

\begin{figure}[h]
\centerline{\raisebox{2.5cm}{$\sigma\ $}
\includegraphics[width=0.4\textwidth]{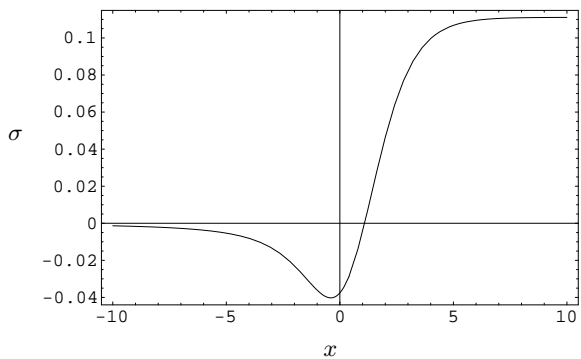}}
\centerline{$\ \ \ \ \ \ \ \ \ x$}
\caption{The ``perturbation'' $\sigma$ for $\beta=-2$.}
\label{eta_deSitt}
\end{figure}

In this model the scale factor is given by
\be
a(\eta )=\ell_0|\eta |^{1+\beta }
\ ,
\label{a_PowLaw}
\ee
with $\beta \le -2$. The case $\beta =-2$ is special, since it corresponds to the de~Sitter space-time with constant Hubble
radius equal to $\ell_0$.
The frequency is given, both for scalar and tensor modes, by
\be
\omega^2(x)=
\left(1+\beta\right)^2\,{\rm e}^{2\,x}-\left(\beta+\frac12\right)^2
\ ,
\label{omega2_PL}
\ee
and the horizon flow functions read
\be
\epsilon_1=\frac{2+\beta }{1+\beta }
\ ,\quad
\epsilon_n = 0
\ ,
\quad
n>1
\ .
\label{eps_PowLaw}
\ee
For the particular case $\beta=-2$, we plot the frequency
in Fig.~\ref{omega2_deSitt}, the function $\xi(x)$ in
Fig.~\ref{xi_deSitt} and the perturbation $\sigma(x)$
in Fig.~\ref{eta_deSitt}.
\par
Eq.~(\ref{osci}) with the initial conditions (\ref{init_cond_on_mu})
can be solved analytically and the exact power spectra
at $x_f\to-\infty$ are given by (see,
e.g.~Refs.~\cite{lythstewart,martin_schwarz,abbott,martin_PL})
\begin{subequations}
\be
&&
\mathcal{P}_{\zeta}=
\frac{1}{\pi\,\epsilon_1\,m_{\rm Pl}^2}\,
\frac{1}{l_0^2}\,f(\beta)\,k^{2\,\beta+4}
\nonumber
\\
\label{P_ST_PawLaw}
\\
&&
\mathcal{P}_h=
\frac{16}{\pi\,m_{\rm Pl}^2}\,
\frac{1}{l_0^2}\,f(\beta)\,k^{2\,\beta+4}
\ ,
\nonumber
\ee
where
\be
f(\beta):=
\frac{1}{\pi}\,\left[
\frac{\Gamma\left(\left|\beta+1/2\right|\right)}
{2^{\beta+1}}\right]^2
\ ,
\label{f_beta}
\ee
\end{subequations}
and $f(\beta=-2)=1$~\footnote{The case $\beta =-2$ is singular
($\epsilon _1=0$ and the expression of the scalar power spectrum
blows up) and should be considered separately.
However, there are no scalar perturbations in De~Sitter space-time
to first order.
}.
The spectral indices and their runnings can also be calculated
from Eqs.~(\ref{n_def}) and (\ref{alpha_def}) and one finds
$n_{\rm S}-1=n_{\rm T}=2\,\beta+4$ and
$\alpha_{\rm S}=\alpha_{\rm T}=0$.
Exact scale-invariance is obtained for $\beta =-2$.
\subsection{Leading order results}
\label{Power-law inflation_leading}
The leading order expressions in the perturbative expansion
of Section~\ref{pert}, as well as the adiabatic expansion
of Section~\ref{adiab}, lead to the same result as in
Ref.~\cite{martin_schwarz},
\begin{subequations}
\be
&&
\mathcal{P}_{\zeta}\simeq\frac{1}{\pi\,\epsilon_1\,
m_{\rm Pl}^2}\,\frac{1}{l_0^2}\,g^{(0)}(\beta)\,k^{2\,\beta+4}
\nonumber
\\
\label{P_ST_PawLaw_0}
\\
&&
\mathcal{P}_h\simeq\frac{16}{\pi\,m_{\rm Pl}^2}\,
\frac{1}{l_0^2}\,g^{(0)}(\beta)\,k^{2\,\beta+4}
\ ,
\nonumber
\ee
where the function $g^{(0)}(\beta)$ is given by
\be
g^{(0)}(\beta):=
\frac{2\,e^{2\,\beta+1}}{\left|2\,\beta+1\right|^{2\,\beta +2}}
\ .
\label{g_0_beta}
\ee
\end{subequations}
This yields a relative error for the amplitude of
the power spectrum
\be
\Delta_P^{(0)}:=
100\,\left|\frac{f(\beta)-g^{(0)}(\beta)}{f(\beta)}\right|
\%
\ ,
\label{b_error}
\ee
which decreases for increasing $|\beta|$, as can be seen from Fig.~\ref{errors1}, but is rather large (about $10\%$)
for the de~Sitter space-time.
The spectral indices and their runnings are instead predicted
exactly as $n_{\rm S}-1=n_{\rm T}=2\,\beta+4$ and
$\alpha_{\rm S}=\alpha_{\rm T}=0$.
\subsection{Next-to-leading order results}
\label{Power-law inflation_next-to-leading}
We shall now compare the corrections to the amplitude
of the power spectra coming from the two different
expansions we described in Section~\ref{schiffmethod}.
\subsubsection{Perturbative expansion}
\begin{table*}[ht]
\centerline{
\begin{tabular}{|c|c|c|c|c|c|c|c|c|c|}
\hline
$\beta$ 
& $-2$ 
& $-3$ 
& $-4$ 
& $-5$ 
& $-6$
& $-7$
& $-8$
& $-9$
& $-10$ 
\\
\hline
$g_{(1)}^{\rm GREEN}$ 
& $+9.6\cdot 10^{-5}$ 
& $+7.3\cdot 10^{-4}$
& $+7.1\cdot 10^{-4}$
& $+6.4\cdot 10^{-4}$
& $+5.8\cdot 10^{-4}$
& $+5.2\cdot 10^{-4}$
& $+4.8\cdot 10^{-4}$
& $+4.4\cdot 10^{-4}$
& $+4.2\cdot 10^{-4}$
\\
\hline
$\Delta_P^{(0)}$ 
& $10.4\%$ 
& $6.4\%$
& $4.6\%$
& $3.6\%$
& $3.0\%$
& $2.5\%$
& $2.2\%$
& $1.9\%$
& $1.7\%$
\\
\hline
$\Delta_P^{(1)}$ 
& $10.4\%$ 
& $6.3\%$
& $4.6\%$
& $3.6\%$
& $2.9\%$
& $2.5\%$
& $2.1\%$
& $1.9\%$
& $1.7\%$
\\
\hline
\end{tabular}
}
\caption{Next-to-leading order improvement for the power spectra
($g_{(1)}^{\rm GREEN}$) and total final error
($\Delta_P^{(1)}$) to first order in $\varepsilon$.
The latter is essentially the same as to leading order
($\Delta_P^{(0)}$).}
\label{improveP}
\end{table*}
\begin{table*}[ht]
\centerline{
\begin{tabular}{|c|c|c|c|c|c|c|c|c|c|}
\hline
$\beta$ 
& $-2$ 
& $-3$ 
& $-4$ 
& $-5$ 
& $-6$
& $-7$
& $-8$
& $-9$
& $-10$
\\
\hline
$\phi_{{\rm I}(1)}$ 
& $+3.2\cdot 10^{-3}$
& $+1.1\cdot 10^{-3}$
& $+6.0\cdot 10^{-4}$
& $+3.6\cdot 10^{-4}$
& $+2.4\cdot 10^{-4}$
& $+1.7\cdot 10^{-4}$
& $+1.3\cdot 10^{-4}$
& $+1.0\cdot 10^{-4}$
& $+8.1\cdot 10^{-5}$
\\
\hline
$\gamma_{{\rm I}(1)}$ 
& $-2.4\cdot 10^{-19}$
& $-5.9\cdot 10^{-20}$
& $-2.6\cdot 10^{-20}$
& $-1.5\cdot 10^{-20}$
& $-9.4\cdot 10^{-21}$
& $-6.6\cdot 10^{-21}$
& $-4.8\cdot 10^{-21}$
& $-3.7\cdot 10^{-21}$
& $-2.9\cdot 10^{-21}$
\\
\hline
$\phi_{{\rm II}(1)}$ 
& $+4.5\cdot 10^{-3}$ 
& $+1.6\cdot 10^{-3}$ 
& $+8.2\cdot 10^{-4}$ 
& $+5.0\cdot 10^{-4}$ 
& $+3.3\cdot 10^{-4}$ 
& $+2.4\cdot 10^{-4}$ 
& $+1.8\cdot 10^{-4}$ 
& $+1.4\cdot 10^{-4}$ 
& $+1.1\cdot 10^{-4}$ 
\\
\hline
$\gamma_{{\rm II}(1)}$ 
& $-3.5\cdot 10^{-2}$ 
& $-1.2\cdot 10^{-2}$ 
& $-6.4\cdot 10^{-3}$ 
& $-3.8\cdot 10^{-3}$ 
& $-2.6\cdot 10^{-3}$ 
& $-1.8\cdot 10^{-3}$ 
& $-1.4\cdot 10^{-3}$ 
& $-1.1\cdot 10^{-3}$ 
& $-8.6\cdot 10^{-4}$ 
\\
\hline
$g_{(1)}^{\rm AD}$ 
& $+1.1\cdot 10^{-1}$
& $+6.3\cdot 10^{-2}$
& $+4.5\cdot 10^{-2}$
& $+3.5\cdot 10^{-2}$
& $+2.9\cdot 10^{-2}$
& $+2.4\cdot 10^{-2}$
& $+2.1\cdot 10^{-2}$
& $+1.8\cdot 10^{-2}$
& $+1.6\cdot 10^{-2}$
\\
\hline
$\Delta_P^{(1)}$ 
& $0.83\%$ 
& $0.50\%$
& $0.35\%$
& $0.27\%$
& $0.22\%$
& $0.18\%$
& $0.16\%$
& $0.14\%$
& $0.12\%$
\\
\hline
\end{tabular}
}
\caption{First order coefficients $\phi_{(1)}$ and $\gamma_{(1)}$,
correction ($g_{(1)}^{\rm AD}$) and total relative error
($\Delta_P^{(1)}$) for the power spectrum
to first order in $\delta$.}
\label{improveA}
\end{table*}
From the next-to-leading expressions in Section~\ref{next-to-leading_results},
we obtain that the relative correction to the power
spectra is given by the function
$g_{(1)}^{\rm GREEN}:=g_{(1)\,\rm S}^{\rm GREEN}
=g_{(1)\,\rm T}^{\rm GREEN}$ in
Eq.~(\ref{g(1)}) evaluated at $x=x_f\ll x_*$.
This quantity can be determined numerically and one
finds that it approaches an asymptotical finite value
for $x_f\to-\infty$.
Some examples are given in Table~\ref{improveP} for
$x_f=-13$, from which it is clear that the improvement
over the leading order is small.
Once the correction has been included, the total relative
error on the power spectra (denoted by $\Delta_P^{(1)}$)
therefore remains of the same order as the leading order
error $\Delta_P^{(0)}$.
\subsubsection{Adiabatic expansion}
\begin{figure}[h]
\centerline{\raisebox{2.5cm}{$\Delta_P\ $}
\includegraphics[width=0.4\textwidth]{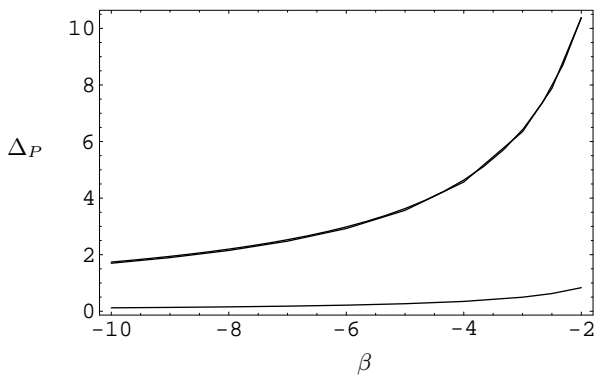}}
\centerline{$\ \ \ \ \ \ \ \ \ \ \ \ \ \ \ \beta$}
\caption{The total error on the power spectrum
to leading order and to next-to-leading order for
the perturbative correction (superposed upper lines) and
to next-to-leading order for the adiabatic correction
(lower line).}
\label{errors1}
\end{figure}
\begin{figure}[t]
\centerline{\includegraphics[width=0.42\textwidth]{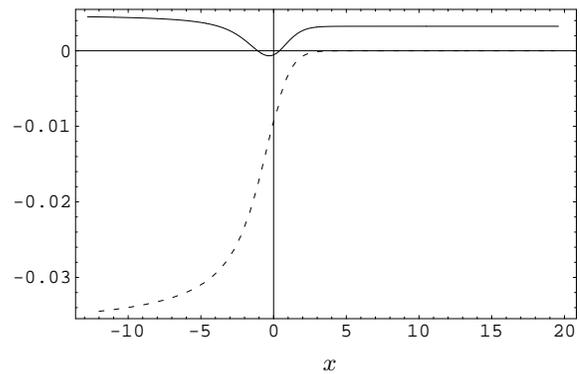}}
\centerline{$\ \ \ \ \ \ \ \ \ x$}
\caption{The functions $\phi_{(1)}(x)$ (solid line) and
$\gamma_{(1)}(x)$ (dotted line) for $\beta=-2$.}
\label{PhiGammaPL}
\end{figure}
We analogously evaluate the coefficients $\phi_{(1)}$
and $\gamma_{(1)}$ numerically.
For the case $\beta=-2$, the result is plotted in
Fig.~\ref{PhiGammaPL}, from which it appears that the
functions $\phi_{(1)}(x)$ and $\gamma_{(1)}(x)$
converge to finite asymptotic values for $x\to\pm\infty$.
Since the same behavior is also found for all $\beta<-2$,
we exhibit the values of such coefficients (with
$x_i=20$ and $x_f=-13$) in Table~\ref{improveA},
together with the corresponding next-to-leading
order relative improvement
[$g_{(1)}^{\rm AD}(x_f)$ as defined in Eq.~(\ref{g_1})]
and the total error on the power spectra ($\Delta_P^{(1)}$). 
On comparing with the last row in Table~\ref{improveP},
we can therefore conclude that the adiabatic corrections
are significantly better than those obtained from the
perturbative expansion and, in fact, yield the amplitudes
with extremely high accuracy (see also Fig.~\ref{errors1}).
In fact, this method seems (at least) as accurate as the
next-to-leading order in the uniform approximation employed in
Ref.~\cite{HHHJM}. 
\section{Conclusions}
\setcounter{equation}{0}
\label{conc}
In this paper we have improved the WKB approximation for the
purpose of estimating the spectra of cosmological perturbations
during inflation.
We found general formulae for amplitudes, spectral indices and
$\alpha$-runnings of the fluctuations to next-to-leading order
both in the adiabatic expansion of Ref.~\cite{langer} and a new
perturbative expansion which makes use of the Green's function
technique.
\par
We have then applied our method to power-law inflation in order
to test it against exact results.
It is known that the spectral indices and their runnings are
obtained exactly in leading order, hence we focussed on the
spectra to next-to-leading order.
We have found that the perturbative corrections remain too small
to yield any significant improvement, whereas the adiabatic
expansion to next-to-leading order reproduces the exact
amplitudes with great accuracy (see Ref.~\cite{HHHJM} for a
similar result with the uniform approximation).
This result does not however mean that the perturbative
expansion will not lead to significant corrections in 
different inflationary scenarios. 
\par
One way of understanding the difference between the two
expansions in the power-law case may be the following.
It is known that the Born~approximation in quantum
mechanics (analogous to our perturbative method with
the Green's function) is good for high multipoles
(angular momenta) $\ell$ of the expansion in spherical
harmonics.
In our approach, the parameter $\beta$ plays the role
of the multipole index $\ell$ for the energy levels of
the hydrogen atom, as can be seen on comparing the
frequency~(\ref{omega2_PL}) with the expression
given in Ref.~\cite{langer}
(see also Ref.~\cite{martin_schwarz}).
The approximation with the Green's function is therefore
expected to yield more significant corrections for large
values of $|\beta|$, which coincide with the regime of
fast-roll.
Indeed, the leading order becomes more and more accurate
for increasing $|\beta|$.
Since the interesting regime for inflation involves small
values of $\beta\sim -2$, it is instead the adiabatic
approximation which seems better since the horizon flow
functions evolve slowly and/or the states are
quasi-classical~\footnote{Cosmological perturbations
are amplified by inflation evolving from vacuum to highly
squeezed states, which resemble classical states in the
amplitude of fluctuations~\cite{polarski}.}.
This may be the reason whereby the adiabatic approximation
works better in the next-to-leading order.
However, in more general cases, either or both methods
may contribute significant corrections.
In this respect, let us remark that the adiabatic
expansion can be straightforwardly applied only for
a linear turning point~\footnote{The same limitation
seems to affect the method used in Ref.~\cite{HHHJM}.},
whereas the perturbative Green's function method is
not so restricted.
\par
Having assessed the accuracy of the method, it is now
natural to use it in order to improve current estimates
for inflationary models which are not exactly solvable.
We are in fact analyzing the slow roll approximation
with our method and will report about it in a future
publication~\cite{new}. 
\acknowledgments
We would like to thank Salman~Habib, Katrin~Heitmann,
Jerome~Martin for discussions and comments on the manuscript.
\end{document}